\documentclass[sigconf]{acmart}

\usepackage{booktabs} 
\usepackage{algorithm}
\usepackage{amsmath}
\usepackage[noend]{algpseudocode}
\usepackage{multirow}
\usepackage{makecell}

\usepackage{tabulary}
\newcolumntype{C}[1]{>{\centering\arraybackslash}p{#1}}
\usepackage{bm}

\copyrightyear{2020} 
\acmYear{2020} 
\setcopyright{acmcopyright}
\acmConference[SIGIR '20]{Proceedings of the 43rd International ACM SIGIR Conference on Research and Development in Information Retrieval}{July 25--30, 2020}{Virtual Event, China}
\acmBooktitle{Proceedings of the 43rd International ACM SIGIR Conference on Research and Development in Information Retrieval (SIGIR '20), July 25--30, 2020, Virtual Event, China}
\acmPrice{15.00}
\acmDOI{10.1145/3397271.3401042}
\acmISBN{978-1-4503-8016-4/20/07}
\begin{document}
\fancyhead{}

\title{Try This Instead: Personalized and Interpretable\\Substitute Recommendation}

\author{Tong Chen}
\affiliation{%
  \institution{\hspace{-0.3cm}The University of Queensland}
}
\email{tong.chen@uq.edu.au}

\author{Hongzhi Yin}
\authornote{Corresponding author; contributing equally with the first author.}
\affiliation{%
  \institution{\hspace{-0.3cm}The University of Queensland}
}
\email{h.yin1@uq.edu.au}

\author{Guanhua Ye}
\affiliation{%
  \institution{\hspace{-0.3cm}The University of Queensland}
}
\email{g.ye@uq.net.au}

\author{Zi Huang}
\affiliation{%
  \institution{\hspace{-0.3cm}The University of Queensland}
}
\email{huang@itee.uq.edu.au}

\author{Yang Wang}
\affiliation{%
  \city{Hefei University of Technology}
}
\email{yangwang@hfut.edu.cn}

\author{Meng Wang}
\affiliation{%
  \city{Hefei University of Technology}
}
\email{eric.mengwang@gmail.com}

\renewcommand{\shorttitle}{Try This Instead: Personalized and Interpretable Substitute Recommendation}

\begin{abstract}
As a fundamental yet significant process in personalized recommendation, candidate generation and suggestion effectively help users spot the most suitable items for them. Consequently, identifying substitutable items that are interchangeable opens up new opportunities to refine the quality of generated candidates. When a user is browsing a specific type of product (e.g., a laptop) to buy, the accurate recommendation of substitutes (e.g., better equipped laptops) can offer the user more suitable options to choose from, thus substantially increasing the chance of a successful purchase. However, existing methods merely treat this problem as mining pairwise item relationships without the consideration of users' personal preferences. Moreover, the substitutable relationships are implicitly identified through the learned latent representations of items, leading to uninterpretable recommendation results. 

In this paper, we propose attribute-aware collaborative filtering (A2CF) to perform substitute recommendation by addressing issues from both personalization and interpretability perspectives. In A2CF, instead of directly modelling user-item interactions, we extract explicit and polarized item attributes from user reviews with sentiment analysis, whereafter the representations of attributes, users, and items are simultaneously learned. Then, by treating attributes as the bridge between users and items, we can thoroughly model the user-item preferences (i.e., personalization) and item-item relationships (i.e., substitution) for recommendation. In addition, A2CF is capable of generating intuitive interpretations by analyzing which attributes a user currently cares the most and comparing the recommended substitutes with her/his currently browsed items at an attribute level. The recommendation effectiveness and interpretation quality of A2CF are further demonstrated via extensive experiments on three real-life datasets.
\end{abstract}

\ccsdesc[500]{Information systems~Recommender systems}

\keywords{Substitute Recommendation; Product Relationship; Interpretability; Collaborative Filtering}

\maketitle

\section{Introduction}\label{sec:intro}
On modern e-commerce platforms, it is a common practice to deploy recommender systems for retrieving items that match users' personal interests \cite{koren2009matrix}. With the increasingly high heterogeneity of user-item interaction data, recommender systems are expected to understand more complex contexts other than user preferences, such as purchase sequences \cite{guo2019streaming,sun2019can}, fine-grained user intents \cite{sun2017collaborative,chen2019air} and social connections \cite{chen2020social}.

More recently, mining product relationships for different online shopping stages have shown its strength in further improving customer satisfaction and sales revenue. One typical line of research is to recommend complementary items that are mutually compatible (e.g., iPhone 11 and phone cases), which is also known as bundle recommendation \cite{pathak2017generating,bai2019personalized}. In a variety of successful attempts, the identified complements can either stimulate further purchases after a user has bought a compatible item \cite{kang2018recommendation,kang2019complete}, or attract potential customers via bundle advertisements beforehand \cite{chen2019matching,zhu2014bundle}. In contrast to complementary relationships, substitutable relationships exist among items that are interchangeable and functionally similar (e.g., iPhone 11 and Samsung Galaxy S10). As a typical decision process in e-commerce \cite{moe2006empirical}, when a user is looking for a particular type of product to buy (e.g., a laptop), she/he tends to first acquire a set of candidate items for comparison, and then pick the most suitable one (if there is any). Correspondingly, in a user's decision-making process, recommending items that are substitutable and even superior to the one currently being browsed can expand the user's view to make a better decision and eventually increase the chance of a successful purchase \cite{zhang2019inferring}. 

\begin{figure}[!t]
\center
\includegraphics[width = 3.3in]{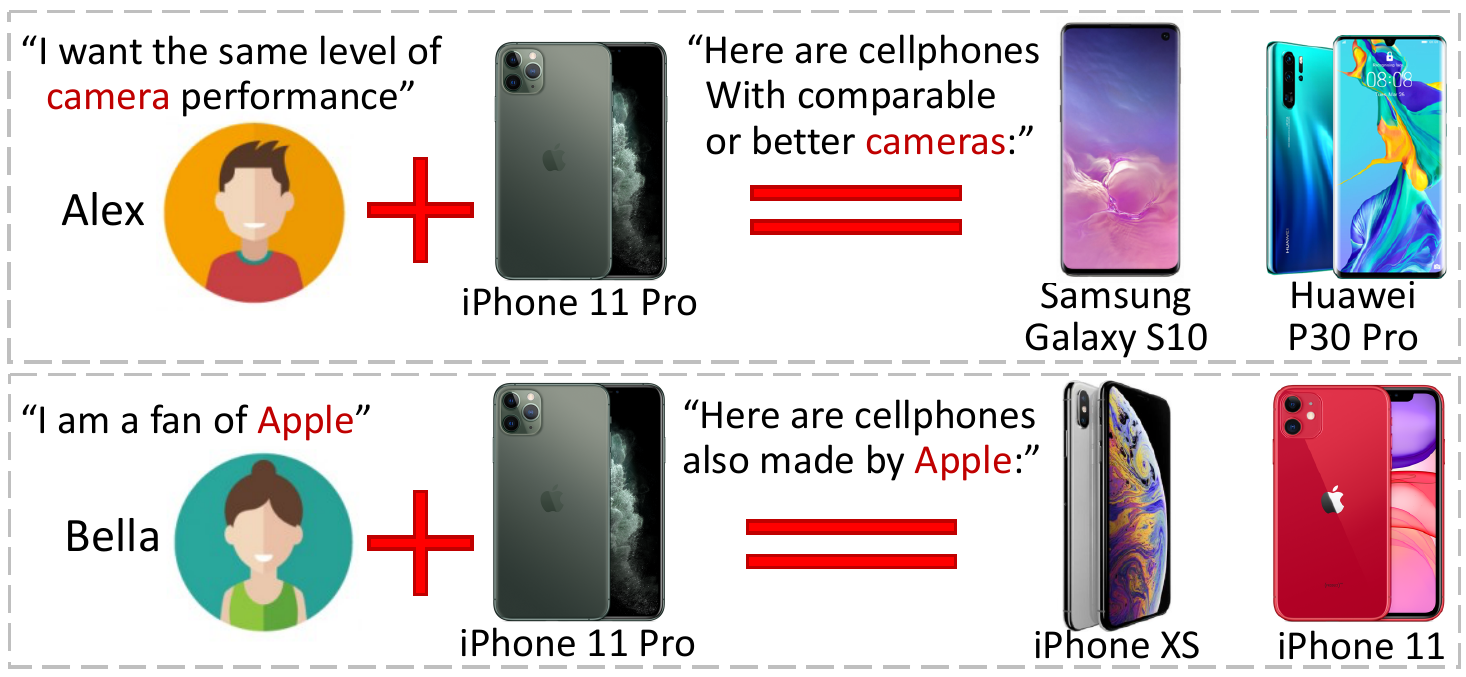}
\vspace{-0.46cm}
\caption{The effect of personalization and interpretability in substitute recommendation.}
\label{Figure:introFig}
\vspace{-0.8cm}
\end{figure}

However, compared with complement recommendation that has been widely studied and applied to multiple domains like fashion recommendation \cite{kang2019complete,song2017neurostylist}, targeted advertising \cite{zhu2014bundle,kouki2019product} and online retail \cite{wan2018representing,bai2019personalized}, the problem of substitute recommendation remains largely unexplored. \cite{mcauley2015inferring} is the first work to systematically investigate product relationships using reviews. In \cite{mcauley2015inferring}, items are modelled via latent Dirichlet allocation (LDA) that captures textual information from reviews, so that the substitutable relationships between any two items can be predicted by comparing their textual contents. With textual features, RRN \cite{zhang2019identifying} and LVA \cite{rakesh2019linked} are two neural network-based extensions that identify substitutable products using feed-forward networks and variational autoencoders respectively. 
 Recent models like PMSC \cite{wang2018path} and SPEM \cite{zhang2019inferring} further leverage structural constraints in observed product co-occurrence (e.g., ``also viewed'') graphs for discriminating substitutes. Despite the importance of generating appropriate substitutable candidates, most existing solutions treat substitute recommendation as a straightforward item-item retrieval task. That is, given an arbitrary item as the query, the recommendation model is expected to output the most relevant items as substitutes based on a pairwise scoring function \cite{mcauley2015inferring,zhang2019inferring,zhang2019identifying}. As a result, the retrieved substitutes are purely conditioned on the query item while different users' personal preferences are neglected. Taking Figure \ref{Figure:introFig} as an example, user Alex and Bella both come across with iPhone 11 Pro (i.e., the query item) when searching for a cellphone to buy. Since Alex cares much about camera quality on cellphones, the generated substitutes should be other top-tier cellphones with comparable or higher camera performance. Meanwhile, Bella simply prefers the Apple brand, so other similarly equipped iPhones are expected to be recommended. Unfortunately, without personalization, all the aforementioned methods will yield the same recommendation results for both users, because they are only determined by the similarity between the query item and substitutes. In \cite{rakesh2019linked}, though the authors extend their proposed LVA to personalized CLVA, CLVA simply introduces user-item matrix factorization as an add-on module to LVA, which is incapable of learning users' explicit demands on fine-grained item attributes (e.g., ``camera'' and ``brand'').

Furthermore, most existing substitute recommendation models are purely built upon latent factor models like word2vec \cite{mikolov2013efficient} and deep neural networks \cite{lecun2015deep}. Though promising recommendation results are reported, a widely recognized drawback of latent factor models is that, the recommendation process is not transparent and the recommendation results are hardly interpretable to users \cite{he2015trirank,zhang2014explicit,zhang2018explainable}. Consequently, it inevitably creates a bottleneck for convincing a user why the recommended substitutes are more suitable than the query item, and how they meet this user's specific needs. 

To this end, in this paper, we aim to advance substitute recommendation by addressing issues from both personalization and interpretability perspectives. On top of user-item interaction records, the growing availability of user reviews offers abundant information about their most cared attributes of a product and their sentiment towards product attributes \cite{he2015trirank,zhang2014explicit,wang2018explainable}. As such, we propose attribute-aware collaborative filtering (A2CF), which makes full use of explicit item attributes to express user-item preferences and item-item substitutions. Instead of directly using extracted attributes as features for item representation learning \cite{mcauley2015inferring,rakesh2019linked,zhang2019identifying}, we decouple user-item interactions into user-attribute and item-attribute interactions, enabling A2CF to explicitly learn a user's attention and an item's performance across various attributes. In this way, we can not only recognize two substitutable items by comparing their attributes, but also integrate users' personal preferences to suggest more suitable substitutes. In addition, as illustrated in Figure \ref{Figure:introFig}, our proposed A2CF also enables personalized explanations. Specifically, we design a novel attribute-aware comparison scheme in A2CF, which infers each user's current demand on specific product attributes, and then highlights these attributes with more advantageous performance provided by the recommended substitutes.

In summary, the contributions of this paper are three-fold:
\begin{itemize}
	\item We identify the shortcomings of existing substitute recommendation methods, and propose a new problem -- personalized and interpretable substitute recommendation that aims to suggest substitutable items customized for users' preferences, along with intuitive explanations.
	\item We propose A2CF, a novel attribute-aware collaborative filtering model. By incorporating explicit user-attribute and item-attribute associations, A2CF simultaneously optimizes both substitution and personalization constraints for recommendation, and further interprets the recommendation results via an attribute comparison scheme.
	\item We conduct extensive experiments on three million-scale datasets. Comparisons with state-of-the-art methods demonstrate the superior recommendation performance of A2CF and its capability to generate high-quality interpretations.
\end{itemize}

\section{Preliminaries}\label{sec:pre}

\subsection{Notations} 
Throughout this paper, all vectors and matrices are respectively denoted by bold lowercase and bold uppercase letters. e.g., user embedding $\textbf{u}$, item embedding $\textbf{v}$, attribute embedding $\textbf{a}$, user-attribute matrix $\textbf{X}$, and item-attribute matrix $\textbf{Y}$. All sets are calligraphic uppercase letters, e.g., user, item, attribute sets are respectively denoted by $\mathcal{U}$, $\mathcal{V}$ and $\mathcal{A}$. All vectors are \textit{\textbf{column vectors}} unless specified, e.g., $\textbf{v} \in \mathbb{R}^{d\times 1}$.

\vspace{-0.2cm}
\subsection{Extracting Attributes and Sentiments}\label{sec:extract_attr}
Attributes are the components, features or properties of an item \cite{he2015trirank}. Based on user reviews on items, attribute-sentiment pairs can be extracted and have shown prominent contributions to review-based recommendation \cite{zhang2014users,zhang2014aspect,wang2018explainable}. As we mainly focus on leveraging the attribute information for recommendation, we apply a well-established tool named Sentires \cite{zhang2014users} to extract attribute-sentiment pairs due to its excellent performance in a wide range of recommendation tasks \cite{zhang2014explicit, guan2019attentive, he2015trirank}. In short, the attribute extraction will result in a sentiment lexicon $\mathcal{Z}$ for the entire review dataset and each entry is $(user, item, attribute, sentiment)$, denoted by $(u,v,a,s)$. The sentiment polarity $s$ is either $+1$ or $-1$, e.g., $(Alex, price, +1)$. To enable subsequent computations, we define the user-attribute matrix $\textbf{X}$ and item-attribute matrix $\textbf{Y}$ as follows. 

\noindent \textbf{Definition 1} (User-Attribute Matrix):
The user-attribute matrix $\textbf{X}$ is a $|\mathcal{U}|\times |\mathcal{A}|$ matrix carrying information on how each user $u_i \in \mathcal{U}$ cares about a specific attribute $a_n \in \mathcal{A}$. Intuitively, a user tends to comment on several attributes (e.g., ``battery'' for electronics and ``taste'' for food) that she/he is most interested in. So, we define each entry $x_{in} \in \textbf{X}$ as:
\vspace{-0.3cm}
\begin{equation}
	x_{in} = \bigg{\{}
	\begin{array}{c}
			\!\!\!0, \,\, \textnormal{if} \,\, u_i \,\, \textnormal{has never mentioned} \,\,a_n \\
			1 + (N-1)\frac{1-\exp(-t_{in})}{1+\exp(-t_{in})}, \,\, \textnormal{otherwise} \\
	\end{array},
\vspace{-0.1cm}
\end{equation}
where $t_{in}$ is the total count of user $u_i$ mentioning attribute $a_n$. $N$ is the scaling factor such that $1 \leq x_{in} \leq N$. Following \cite{zhang2014explicit,wang2018explainable}, we set $N$ as the highest score of the rating system (e.g., 5 for Amazon).

\noindent \textbf{Definition 2} (Item-Attribute Matrix): Similarly, the $|\mathcal{V}| \times |\mathcal{A}|$ item-attribute matrix $\textbf{Y}$ reflects the aggregated feedback of all users on each attribute $a_n \in \mathcal{A}$ of item $v_j \in \mathcal{V}$, where each element $y_{jn}$ is:
\vspace{-0.2cm}
\begin{equation}
	y_{jn} = \bigg{\{}
	\begin{array}{c}
			0, \,\, \textnormal{if} \,\, v_j \,\, \textnormal{has no reviews mentioning} \,\,a_n \\
			\hspace{-1.3cm} 1 + \frac{(N-1)}{1+\exp(-t_{jn} \overline{s}_{jn})}, \,\, \textnormal{otherwise} \\
	\end{array},
\vspace{-0.1cm}
\end{equation} 
where we also let $1 \leq y_{jn} \leq N$, and $t_{jn}$ means that the attribute $a_n$ of item $v_j$ is mentioned $n$ times by all users, while $\overline{s}_{jn}$ is the mean of all the sentiment scores in those $t_{jn}$ mentions.

\section{Attribute-Aware Collaborative Filtering}\label{sec:A2CF}
In matrix $\textbf{X}$, each non-zero entry $x_{in}$ reflects user $u_i$'s amount of attention paid to a specific attribute $a_n$ during shopping. Similarly, for each $y_{jn} \in \textbf{Y}$ where $y_{jn} \neq 0$, it indicates the explicit user feedback on item $v_j$'s attribute $a_n$, which is crowdsourced and summarized from all user reviews. 
We hereby present the design of our proposed attribute-aware collaborative filtering (A2CF) model. 

\vspace{-0.1cm}
\subsection{User-Attribute Collaborative Filtering}
In traditional collaborative filtering (CF) like matrix factorization \cite{koren2009matrix}, given a user-item matrix with partially observed feedback, a fundamental mechanism is to learn the vectorized representations (a.k.a. embedding vectors) of both users and items using available user-item interactions, and then the unobserved values of user-item interactions can be easily estimated via the inner product of two vectors. In A2CF, with the user-attribute matrix $\textbf{X}$, we aim to learn the representations of each user $u_i$ and attribute $a_n$ so that every scalar $x_{in}\in \textbf{{X}}$ (i.e., the pairwise user-attribute preference) can be inferred. Specifically, we achieve this goal by minimizing the following squared loss function:
\vspace{-0.1cm}
\begin{equation}\label{eq:loss_UA}
	L_{UA} = \sum_{x_{in} \in \mathcal{D}_{UA}}(x_{in} - \widehat{x}_{in})^2,
\vspace{-0.1cm}
\end{equation}
where $\mathcal{D}_{UA} = \{x_{in} \mid x_{in}\in\textbf{X} \land x_{in}\neq 0\}$ is the set of all observed user-attribute relationships, and $\widehat{x}_{in}$ is the estimated value between user $u_i$ and attribute $a_n$. In what follows, we describe the computation of the interaction score $\widehat{x}_{in}$ in detail.

With the given $u_i$ and $a_n$, we firstly model the nonlinear pairwise interaction between them. As pointed out by \cite{he2017neuralcol,wang2019neural}, traditional matrix factorization-based CF methods are subject to limited expressiveness because pairwise interactions are linearly modelled via two vectors' inner product. Thus, in A2CF, we leverage a residual feed-forward network with $l$ hidden layers for modelling pairwise user-attribute interactions:
\vspace{-0.1cm}
\begin{equation}\label{eq:ffn}
	\textbf{h}_{l'} =  \textbf{h}_{l'-1} + \textnormal{ReLU}(\textbf{W}_{l'}\textbf{h}_{l'-1} + \textbf{b}_{l'}),
\vspace{-0.1cm}
\end{equation}
where $l'\leq l$, ReLU is the rectified linear unit for nonlinear activation, while $\textbf{h} \!\!\in\! \mathbb{R}^{d\times 1}$, $\textbf{W} \!\!\in\! \mathbb{R}^{d\times d}$ and $\textbf{b} \!\!\in\! \mathbb{R}^{d\times 1}$ are respectively the latent representation, weight and bias in each layer. $\textbf{h}_0$ is initialized by concatenating both user and attribute embeddings:
\vspace{-0.2cm}
\begin{equation}
	\textbf{h}_0 = [\textbf{u}_i; \textbf{a}_n],
\vspace{-0.1cm}
\end{equation}
where $\textbf{u}_i \in \mathbb{R}^{d \times 1}$ and $\textbf{a}_n \in \mathbb{R}^{d \times 1}$ are embeddings of user $u_i$ and attributes $a_n$ to be learned. 

On top of the $l$-layer residual network, we utilize the final latent representation $\textbf{h}_l$ to compute a scalar output $\widehat{x}_{in}$ as the estimated score for the given user-attribute pair: 
\vspace{-0.1cm}
\begin{equation}
	\widehat{x}_{in} = \tanh_{[1,N]}(\textbf{w}_x^{\top} \textbf{h}_l),
\vspace{-0.1cm}
\end{equation}
where $\tanh_{[1,N]}(\cdot)$ is our modified version of $\tanh(\cdot)$ function to emit prediction results in a customized range. Specifically, for an arbitrary scalar input $r$, the computation process of $\tanh_{[1,N]}(r)$ is:
\vspace{-0.4cm} 
\begin{align}
	\tanh_{[1,N]}(r)
	&= \frac{N-1}{2}\cdot \frac{\exp(2r)-1}{\exp(2r)+1} + \frac{N+1}{2} \nonumber \\
	&= \frac{N\exp(2r)+1}{\exp(2r)+1}.
\vspace{-1cm}
\end{align}
As $\tanh_{[1,N]}(r) \approx 1$ when $r = -\infty$ and $\tanh_{[1,N]}(r) \approx N$ when $r = +\infty$, function $\tanh_{[1,N]}(\cdot)$ forces $\widehat{x}_{in}$ to fall in the rage of $[1,N]$.

So far, we can summarize the computation of $\widehat{x}_{in}$ as:
\vspace{-0.1cm}
\begin{equation}\label{eq:x_predict}
	\widehat{x}_{in} = \tanh_{[1,N]} \big{(}\textbf{w}^{\top}_{x} H_l^{UA}([\textbf{u}_i;\textbf{a}_n])\big{)},
\vspace{-0.1cm}
\end{equation}
where we use $H^{UA}_l(\cdot)$ to denote the $l$-layer residual feed-forward network in Eq.(\ref{eq:ffn}), and $\textbf{w}_{x} \in \mathbb{R}^{d\times 1}$ is the projection weight. With $\widehat{x}_{in}$ computed, we are now able to optimize the learned user and attribute representations by minimizing the loss in Eq.(\ref{eq:loss_UA}).

\vspace{-0.1cm}	
\subsection{Item-Attribute Collaborative Filtering}
Similar to $\textbf{X}$, given the item-attribute matrix $\textbf{Y}$, we can also learn the representations of item $v_j$ and attribute $a_n$. With an estimation of the item-attribute relationship $\widehat{y}_{jn}$ (i.e., the quality of attribute $a_n$ on item $v_j$), we define another squared loss as the following:
\vspace{-0.1cm}
\begin{equation}\label{eq:loss_IA}
	L_{IA} = \sum_{y_{jn} \in \mathcal{D}_{IA}}(y_{jn} - \widehat{y}_{jn})^2,
\vspace{-0.1cm}
\end{equation}
with $\mathcal{D}_{IA} = \{y_{in} \mid y_{in}\in\textbf{Y} \land y_{in}\neq 0\}$ containing all non-zero entries in $\textbf{Y}$. We hereby adopt a similar approach as described for user-attribute collaborative filtering to predict $\widehat{y}_{jn}$:
\vspace{-0.1cm}
\begin{equation}\label{eq:y_predict}
	\widehat{y}_{in} = \tanh_{[1,N]} \big{(}\textbf{w}^{\top}_{y} H_l^{IA}([\textbf{v}_j;\textbf{a}_n])\big{)},
\vspace{-0.1cm}
\end{equation}
where the rescaled activation function $\tanh_{[1,N]}(\cdot)$ is also adopted, and $\textbf{w}_y \in \mathbb{R}^{d\times1}$ is the weight vector. $H^{IA}_{l}(\cdot)$ is another $l$-layer residual feed-forward network whose input is the concatenation of item embedding $\textbf{v}_j \in \mathbb{R}^{d\times1}$ and attribute embedding $\textbf{a}_n \in \mathbb{R}^{d\times1}$. Note that $H^{IA}_{l}(\cdot)$ shares the same architecture with $H^{UA}_l(\cdot)$ but uses a different set of network weights and biases. 

At the same time, since the learned attributes will serve the purpose of quantifying both user preferences and item characteristics, we make all attribute embeddings shared between user-attribute CF and item-attribute CF. Hence, this enables the joint optimization of both objectives, leading to stronger expressiveness of learned attribute representations.

\vspace{-0.2cm}
\subsection{Estimating Missing Entries in $\textbf{X}$ and $\textbf{Y}$}\label{sec:estimate}
Suppose that we have already learned the representations for users, items and attributes that yield sufficiently small values for both $L_{UA}$ and $L_{IA}$. Before entering the next stage where all the learned representations are further fine-tuned for recommendation, we need to infer the unobserved user-attribute and item-attribute values. In other words, we estimate the missing entries (i.e., 0-valued elements) in $\textbf{X}$ and $\textbf{Y}$ with Eq.(\ref{eq:x_predict}) and Eq.(\ref{eq:y_predict}) respectively, which can be viewed as two well-trained nonlinear regressors. Formally, this process is written as:
\vspace{-0.2cm}
\begin{equation}\label{eq:est_X_Y}
	\widetilde{x}_{in} = \bigg{\{}
	\begin{array}{c}
			\widehat{x}_{in}, \,\, \textnormal{if} \,\, x_{in} = 0 \\
			x_{in}, \,\, \textnormal{if} \,\, x_{in} \neq 0 \\
	\end{array}\!\!, \,\,	
	\widetilde{y}_{jn} = \bigg{\{}
	\begin{array}{c}
			\widehat{y}_{jn}, \,\, \textnormal{if} \,\, y_{jn} = 0 \\
			y_{jn}, \,\, \textnormal{if} \,\, y_{jn} \neq 0 \\
	\end{array}\!\!.
\vspace{-0.1cm}
\end{equation}

We use $\widetilde{\textbf{X}}$ and $\widetilde{\textbf{Y}}$ to denote the fully estimated user-attribute and item-attribute matrices. In addition, we use $\widetilde{\textbf{x}}_i =[\widetilde{x}_{i1}, \widetilde{x}_{i2}, \cdots, \widetilde{x}_{i|\mathcal{A}|}]$ to denote the $i$-th row of $\widetilde{\textbf{X}}$. Specifically, each element in $\widetilde{\textbf{x}}_i$ corresponds to user $u_i$'s preference on an item attribute, parameterized in the range of $[1,N]$. In a similar way, $\widetilde{\textbf{y}}_j$ refers to the $j$-th row of $\widetilde{\textbf{Y}}$, which is actually a collection of item $v_j$'s performance scores (within $[1,N]$ range) regarding all attributes.

\vspace{-0.2cm}
\subsection{Uniting Substitution and Personalization Constraints for Recommendation}
Given user $u_i$ and query item $v_q$ that she/he is currently browsing, we would like to recommend a list of items that are: (1) strong alternatives to $v_q$; and (2) of great interest to $u_i$. To achieve this, we extend Bayesian personalized ranking (BPR) \cite{rendle2009bpr} to the substitute recommendation setting. Despite being widely adopted for personalized recommendation, BPR is originally designed for optimizing pairwise user-item scoring functions, and lacks the consideration of affinity between two substitutable items. In this section, we propose a variant of BPR to unify both substitution and personalization. For user $u_i$ and query item $v_q$, we firstly pair them with the ground truth substitute $v_{j^+}$, denoted by $(u_i, v_q, v_{j^+})$, implying that user $u_i$ has eventually chosen $v_{j^+}$ after browsing $v_q$. Correspondingly, a negative case $(u_i, v_q, v_{j^-})$ can be constructed, where $v_{j^-}$ holds either or both of the properties: (1) $v_{j^-}$ is never visited by $u_i$; (2) $v_{j^-}$ is not substitutable to $v_q$. Thus, we can form a set of training samples consisting of $(u_i, v_q, v_{j^+}, v_{j^-})$, denoted by $\mathcal{D}_{Rec}$. With the training samples, we aim to learn a triplet ranking function $f(\cdot,\cdot,\cdot)$ to generate scalar ranking scores such that $f(u_i, v_q, v_{j^+}) > f(u_i, v_q, v_{j^-})$. In this regard, we define the following loss function that integrates BPR with substitutable item relationships, which we term $L_{BPR-S}$:
\vspace{-0.1cm}
\begin{align}\label{eq:BPR-S}
	&\!\!\!\!L_{BPR-S} = - \log \!\!\!\!\!\!\!\!\! \prod_{(u_i, v_q, v_{j^+}, {v_{j^-}})\in \mathcal{D}_{Rec}} \!\!\!\!\!\!\! \sigma \Big{(}f(u_i, v_q, v_{j^+}) - f(u_i, v_q, v_{j^-})\Big{)} \nonumber\\
	&\!\!\!\!= - \!\!\!\!\!\!\!\!\!\! \sum_{(u_i, v_q, v_{j^+}, v_{j^-})\in \mathcal{D}_{Rec}} \!\!\!\!\!\!\!\!\! {\log \bigg{(}\sigma \Big{(}f(u_i, v_q, v_{j^+}) - f(u_i, v_q, v_{j^-})\Big{)} \bigg{)}},
\vspace{-0.3cm}
\end{align}
where $\sigma(\cdot)$ is the $Sigmoid$ function. In general, we expect $f(\cdot,\cdot,\cdot)$ to fulfill two special constraints, namely the \textit{\textbf{substitution constraint}} and \textit{\textbf{personalization constraint}}, to produce accurate recommendations. Before explaining the details, we let $f(u_i, v_q, v_j) = \gamma f_S(v_q,v_j) + (1-\gamma)f_P(u_i, v_j)$, and rewrite Eq.(\ref{eq:BPR-S}) as follows:
\vspace{-0.1cm}
\begin{align}\label{eq:BPR-S-rewrite}
	 L_{BPR-S} = & - \!\!\!\!\!\!\!\!\!\! \sum_{(u_i, v_q, v_{j^+}, v_{j^-})\in \mathcal{D}_{Rec}} \!\!\!\!\!\!\!\!\!\! \sigma' \Big{(}\gamma f_S(v_q, v_{j^+}) + (1-\gamma)f_P(u_i, v_{j^+})  \nonumber\\
	& - \gamma f_S(v_q, v_{j^-}) - (1-\gamma) f_P(u_i, v_{j^-}) \Big{)},
\vspace{-0.6cm}
\end{align}
where we let $\sigma'(\cdot) = \log(\sigma(\cdot))$ to be succinct, $f_S(\cdot,\cdot)$ and $f_P(\cdot,\cdot)$ are respectively the pairwise item-item and user-item scoring functions complying with the substitution and personalization constraints. $\gamma \in (0,1)$ is a hyperparameter for adjusting the trade-off between two terms. Next, we introduce the designs of $f_S(\cdot,\cdot)$ and $f_P(\cdot,\cdot)$.

\textbf{Implementing Substitution Constraint.} For an arbitrary item $v_j$ ($\, j\in\{j^+,j^-\}$) paired with the query item $v_q$, $f_S(\cdot,\cdot)$ is expected to generate a high affinity score $f_S(v_q, v_j)$ if they are mutually substitutable, and a relatively low score if not. Different from substitute recommenders that calculate item-item similarities in a pure latent factor-based manner \cite{zhang2019inferring,rakesh2019linked,wang2018path}, $f_S(\cdot,\cdot)$ additionally incorporates information of explicit attributes on top of the learned item representations. Intuitively, a substitutable item must be functionally similar to the query item $v_q$, which is reflected by their similar distributions over attributes. That is to say, we can compare two items via their attribute information. So, item $v_j$ can be represented as a weighted combination of all attribute embeddings:
\vspace{-0.2cm}
\begin{equation}
	\widetilde{\textbf{v}}_{j} = \sum_{n=1}^{|\mathcal{A}|} \varphi^{qj}_{n}\textbf{a}_{n},
\vspace{-0.1cm}
\end{equation}
where $\widetilde{\textbf{v}}_{j}$ is the representation of item $v_j$ after the attribute aggregation, $\{\varphi^{qj}_{n}\}_{n=1}^{|\mathcal{A}|}$ is a probability distribution (i.e., $\sum_{n=1}^{|\mathcal{A}|} \varphi^{qj}_{n} = 1$) over all attributes, which is dependent on $v_q$ and $v_j$:
\vspace{-0.1cm}
\begin{equation}\label{eq:softmax1}
	\{\varphi^{qj}_n\}_{n=1}^{|\mathcal{A}|} = \textnormal{softmax}(\frac{\widetilde{\textbf{y}}_{q} \odot \widetilde{\textbf{y}}_{j}}{\beta}),
\vspace{-0.1cm}
\end{equation}
where $\widetilde{\textbf{y}}_q$ and $\widetilde{\textbf{y}}_j$ are respectively the $q$-th and $j$-th row of the estimated item-attribute matrix $\widetilde{\textbf{Y}}$, $\odot$ is the element-wise product of two vectors, and $\beta$ is the scaling factor that controls the strength of dominating attributes. Instead of picking only a small fraction of most relevant attributes to represent each item \cite{zhang2014explicit,mcauley2015inferring}, we use $softmax$ to avoid potential information loss by discriminatively considering all attributes. As discussed in Section \ref{sec:estimate}, in $\widetilde{\textbf{y}}_q$ and $\widetilde{\textbf{y}}_j$, the $n$-th elements $\widetilde{y}_{qn}$ and $\widetilde{y}_{jn}$ reflect both items' quality on attribute $a_n$. Apparently, if item $v_j$ is similar to the query item $v_q$ on attribute $a_n$, Eq.(\ref{eq:softmax1}) will generate a large weight $\varphi^{qj}_{n}$. The output $\{\varphi^{qj}_{n}\}_{n=1}^{|\mathcal{A}|}$ essentially carries all attribute-wise similarity scores between $v_q$ and $v_j$. By setting $\beta$ to different values, we can amplify or weaken the influence of larger values in $\{\varphi^{qj}_{n}\}_{n=1}^{|\mathcal{A}|}$. 

Then, we compute the substitution affinity score between $v_q$ and $v_j$:
\vspace{-0.1cm}
\begin{equation}\label{eq:projection}
	f_S(v_q, v_j) = \textbf{w}_s^{\top}[\textbf{v}_q \odot \textbf{v}_j; \widetilde{\textbf{v}}_{j}],
\vspace{-0.02cm}
\end{equation}
where $\textbf{w}_s \in \mathbb{R}^{2d\times 1}$ is the weight for linear projection, and $\textbf{v}_q \odot \textbf{v}_j$ models the element-wise similarity between the latent vectors of two items. In Eq.(\ref{eq:projection}), we concatenate information from two aspects: (1) the comparison between the latent representations of $v_q$ and $v_j$; (2) the attribute-aware representation of $v_j$ generated by comparing explicit item attributes. As this provides substantial information for generating a discriminative ranking score $f_S(v_q, v_j)$, we adopt a simple linear projection rather than sophisticated deep neural network-based regressors \cite{zhang2019identifying} for efficient computation. 

\textbf{Implementing Personalization Constraint.} For user $u_i$ and item $v_j$, The pairwise scoring function $f_P(u_i, v_j)$ should yield a large score if $v_j$ has been visited by $u_i$, and vice versa. Similar to $f_S(\cdot,\cdot)$, we firstly generate another attribute-aware representation of $v_j$, denoted by $\widetilde{\textbf{v}}_{j}'$:
\vspace{-0.3cm}
\begin{equation}\label{eq:weighted_sum_pers}
	\widetilde{\textbf{v}}_{j}{\!\!'} = \sum_{n=1}^{|\mathcal{A}|} \lambda^{ij}_{n}\textbf{a}_{n},
\vspace{-0.1cm}
\end{equation}
with the attentive weights $\{\lambda^{ij}_{n}\}_{n=1}^{|\mathcal{A}|}$ computed by comparing user $u_i$'s interests on all attributes (i.e., $\widetilde{\textbf{x}}_{i}$) and item $v_j$'s performance on all attributes (i.e., $\widetilde{\textbf{y}}_{j}$):
\vspace{-0.2cm}
\begin{equation}\label{eq:softmax2}
	\{\lambda^{ij}_{n}\}_{n=1}^{|\mathcal{A}|} = \textnormal{softmax}(\frac{\widetilde{\textbf{x}}_{i} \odot \widetilde{\textbf{y}}_{j}}{\epsilon}),
\vspace{-0.1cm}
\end{equation}
where we also adopt a scaling factor $\epsilon$ to either diverge or smoothen the probability distribution, and $\widetilde{\textbf{x}}_{i} \in \widetilde{\textbf{X}}$ quantifies user $u_i$'s concerns about each attribute. Intuitively, Eq.(\ref{eq:softmax2}) justifies how well item $v_j$'s attribute quality (i.e., $\widetilde{\textbf{y}}_{j}$) aligns with $u_i$'s demand, and then Eq.(\ref{eq:weighted_sum_pers}) generates a combinatorial representation for item $v_j$ as a weighted sum of all different attributes. Afterwards, a lightweight linear projection scheme is applied to generate the personalization ranking score:
\vspace{-0.1cm}
\begin{equation}\label{eq:projection2}
	f_P(u_i, v_j) = \textbf{w}_p^{\top}[\textbf{u}_i \odot \textbf{v}_j; \widetilde{\textbf{v}}_{j}{\!\!'}],
\vspace{-0.2cm}
\end{equation}
where $\textbf{w}_p \in \mathbb{R}^{2d\times 1}$ is the projection weight, and $\textbf{u}_i \odot \textbf{v}_j$ further infuses the comparison between the user's and item's latent representations to allow for precise user-item preference prediction.

\vspace{-0.2cm}
\subsection{Optimization Strategy}\label{sec:opt}
As A2CF is built upon the deep neural network structure, we learn the model parameters on multiple objectives with a mini-batch stochastic gradient decent algorithm, namely Adam \cite{kingma2014adam} optimizer. To be specific, we firstly optimize $L' = L_{UA} + L_{IA}$, i.e., the combined loss for user-attribute CF and item-attribute CF for $T_1$ iterations. Then, we estimate $\widetilde{\textbf{X}}$ and $\widetilde{\textbf{Y}}$ matrices, and optimize the recommendation loss $L_{BPR-S}$ for $T_2$ iterations. Notably, $\widetilde{\textbf{X}}$ and $\widetilde{\textbf{Y}}$ are pre-computed before training the recommendation part, thus avoiding redundant and inefficient computations. 
We update all model parameters in each iteration and repeat the entire training process described above until $L_{BPR-S}$ converges or is sufficiently small. 

We tune the hyperparameters using grid search. Specifically, the latent dimension $d$ is searched in $\{16,32,64,128,256\}$; the depth of the residual feed-forward network $l$ is searched in $\{1,2,3,4,5\}$; the trade-off coefficient $\gamma$ is searched in $\{0.1, 0.3, 0.5, 0.7, 0.9\}$; and both scaling factors $\beta$ and $\epsilon$ are searched in $\{1, d^{0.25}, d^{0.5}, d^{0.75}, d\}$. We will further discuss the impact of these key hyperparameters to the recommendation performance in Section \ref{sec:paramterimpact}. For optimizing the final recommendation loss $L_{BPR-S}$, we draw 5 negative samples $(u_i, v_q, v_{j^-})$ for each positive label $(u_i, v_q, v_{j^+})$ in training set $\mathcal{D}_{Rec}$. We set the the learning rate to $1\!\times\! 10^{-3}$ and batch size to $256$ according to device capacity. 
 To prevent overfitting, we adopt a dropout \cite{srivastava2014dropout} ratio of $0.4$ on all deep layers of A2CF. 
 
\vspace{-0.2cm}
\section{Utilizing A2CF for Recommendation and Interpretation}
After obtaining a fully-trained A2CF model, here we showcase the roadmap towards personalized and interpretable substitute recommendation using A2CF. 

\vspace{-0.2cm}
\subsection{Personalized Substitute Recommendation}
As a natural extension to personalized top-$K$ recommendation \cite{wang2016spore,he2017neuralcol}, the essence of personalized substitute recommendation is to assign a ranking score to each possible item $v_j$ given the context $(u_i, v_q)$ (i.e., the user and query item), and then recommend $K$ top-ranked items to the user. With an input triplet denoted by $(u_i, v_q, v_j)$, recall that we define the triplet ranking function as $f(u_i, v_q, v_j) = \gamma f_S(v_q, v_j) + (1-\gamma) f_P(u_i, v_j)$. Hence, for each $v_j \in \mathcal{V}$, we can generate a ranking score for triplet $(u_i, v_q, v_j)$ by measuring how well the item $v_j$ fits both the substitution and personalization constraints. Ideally, only the items that satisfy both parts will be ranked at the top of the recommendation list.

\vspace{-0.2cm}
\subsection{Personalized Interpretation}
Latent factor-based substitute recommenders \cite{wang2018path,rakesh2019linked,zhang2019inferring} are uninterpretable as the ranking scores are directly computed via the pure latent representations of items and/or users. Consequently, when performing personalized substitute recommendation, a user may find it difficult to understand the ``try this instead'' suggestions from the recommender systems. In contrast, our proposed A2CF is advantageous owing to its capability of generating intuitive interpretations. Specifically, A2CF composes interpretations by considering two aspects: (1) the key attributes $u_i$ will consider when searching for items similar to $v_q$; (2) how the recommended item $v_j$ performs compared with the query item $v_j$ on these key attributes. With the estimated user-attribute matrix $\widetilde{\textbf{X}}$ and item-attribute matrix $\widetilde{\textbf{Y}}$, we address these two points by comparing $v_q$ with $v_j$ in a quantitative way:
\vspace{-0.2cm}
\begin{align}\label{eq:interpretation}
	\{\Delta^{iqj}_n\}_{n=1}^{|\mathcal{A}|} & = \widetilde{\textbf{x}}_i \odot \widetilde{\textbf{y}}_j - \widetilde{\textbf{x}}_i \odot \widetilde{\textbf{y}}_q \nonumber\\
	& = \widetilde{\textbf{x}}_i \odot (\widetilde{\textbf{y}}_j - \widetilde{\textbf{y}}_q).
\vspace{-0.3cm}
\end{align}
In short, Eq.(\ref{eq:interpretation}) aligns the user's demand with each item's quality attribute-wise, and uses subtraction to compute the advantages (or disadvantages) of $v_j$ against $v_q$ regarding each attribute $a_n$, denoted by $\Delta^{iqj}_{n}$. Next, $Z$ attributes $\{a_{n_1}, a_{n_2},...,a_{n_Z}\}$ with highest values in $\{\Delta^{iqj}_n\}_{n=1}^{|\mathcal{A}|}$ are selected to generate the personalized interpretation. In this paper, we leverage a template-based interpretation scheme to state the key reasons for recommending $v_j$ as the substitute of $v_q$ for user $u_i$. The template is as follows:
\vspace{-0.1cm}
\begin{equation}
\!\!\!\!\!\! \left [
	\begin{array}{l}
	\!\textnormal{``Based on the item $v_q$ you are currently browsing, we}\\
		\!\textnormal{recommend you to try $v_j$ instead because it comes with:}\\
	\!\textnormal{[adjective] $a_{n_1}$, [adjective] $a_{n_2}$, $\cdots$, and [adjective] $a_{n_Z}$.''} \\
	\end{array}
\!\!\! \right ]
\end{equation} 
where for $1 \leq z \leq Z$, each $\textnormal{[adjective]}$ is determined via:
\vspace{-0.1cm}
\begin{equation}
	\textnormal{[adjective]} = \bigg{\{}
	\begin{array}{c}
			\hspace{0.7cm} \textnormal{``better''}, \,\, \textnormal{if} \,\, \Delta^{iqj}_{n_{z}} > 0\\
		 \!\!\! \textnormal{``comparable''}, \,\, \textnormal{otherwise} \\
	\end{array}.
\vspace{-0.1cm}
\end{equation}

It is worth mentioning that we prefer assigning a small value to $Z$ (usually below $5$) to keep the interpretations concise, because users have very limited attention and time to read the generated explanations, and it rarely happens for $\Delta^{iqj}_{n_{z}} \leq 0$ when only the top attributes are selected.   

\vspace{-0.2cm}
\section{Experimental Settings}\label{sec:exp_settings}
\subsection{Datasets}
\vspace{-0.1cm}
We consider three real-world review datasets commonly used for recommendation tasks, which are originally crawled from Amazon and made public by \cite{mcauley2015inferring}, and different product categories are treated as separate datasets. We use three of the largest categories, namely Cellphone and Accessories (Cellphone for short), Automotive, and Office. Based on well-established rules adopted in the substitute recommendation literature \cite{mcauley2015inferring,zhang2019identifying,rakesh2019linked,zhang2019inferring}, item substitution relationships are extracted from all datasets. Specifically, two items $v$ and $v'$ are substitutable if: (1) users viewed $v$ also viewed $v'$, or (2) users viewed $v$ eventually bought $v'$. 

Following \cite{li2016point,rendle2009bpr}, we filter out inactive users with less than 5 interacted items and unpopular items visited by less than 5 users. The primary statistics are shown in Table~\ref{table:Dataset}, where each review corresponds to a user-item interaction record. Note that we use the workflow described in Section \ref{sec:extract_attr} to extract attributes from each individual dataset, and attributes with less than 2 mentions are discarded to reduce excessive noise from the data. We split all datasets with the ratio of $80\%$, $10\%$ and $10\%$ for training, validation and test, respectively. 

\begin{table}[!t]
\caption{Statistics of datasets in use.}
\vspace{-0.5cm}
\renewcommand{\arraystretch}{0.9}
\setlength\tabcolsep{3.5pt}
\center
  \begin{tabular}{c c c c c}
    \toprule
    Dataset & \#Review & \#User & \#Item & \#Attribute\\
    \hline
    Cellphone & 1,639,166 & 665,900 & 48,763 & 2,178 \\
	Automotive & 2,212,111 & 420,394 & 132,467 & 3,501 \\
	Office & 1,641,901 & 515,813 & 58,534 & 2,801 \\	
    \bottomrule
\end{tabular}
\label{table:Dataset}
\vspace{-0.7cm}
\end{table}

\vspace{-0.1cm}
\subsection{Evaluation Protocols}\label{sec:eva_protocol}

\begin{table*}[!t]
\caption{Recommendation results. Numbers in bold face are the best results for corresponding metrics.}
\vspace{-0.4cm}
\centering
\renewcommand{\arraystretch}{0.9}
\setlength\tabcolsep{7pt}
  \begin{tabular}{|c|c|c|c|c|c||c|c|c|c|}
    \hline
    \multirow{2}{*}{Dataset} & \multirow{2}{*}{Method} & \multicolumn{4}{c||}{HR$@$K}  & \multicolumn{4}{c|}{NDCG$@$K} \\
    \cline{3-10}
    & & K=5 & K=10 & K=20 & K=50 & K=5 & K=10 & K=20 & K=50 \\
    \hline
    \multirow{5}{*}{Cellphone} & Sceptre \cite{mcauley2015inferring} & 0.2009 & 0.2689 & 0.3486 & 0.4605 & 0.1570 & 0.1695 & 0.1998 & 0.2421 \\
    & PMSC \cite{wang2018path} & 0.1085 & 0.1276 & 0.1731 & 0.2023 & 0.0827 & 0.0995 & 0.1086 & 0.1125 \\
    & SPEM \cite{zhang2019inferring} & 0.1187 & 0.1449 & 0.1858 & 0.2251 & 0.0951 & 0.1034 & 0.1135 & 0.1300 \\
    & CLVA \cite{rakesh2019linked} & 0.1795 & 0.2621 & 0.3434 & 0.5167 & 0.1224 & 0.1421 & 0.1729 & 0.1976 \\
    \cline{2-10}
    & \textbf{A2CF}& \textbf{0.2466} & \textbf{0.3449} & \textbf{0.4548} & \textbf{0.6229} & \textbf{0.1663} & \textbf{0.1980} & \textbf{0.2263} & \textbf{0.2845} \\
     \hline
     \hline
    \multirow{5}{*}{Automotive} & Sceptre \cite{mcauley2015inferring} & 0.1618 & 0.1709 &  0.2441 & 0.3346 & 0.1021 & 0.1087 & 0.1225 & 0.1401 \\
    & PMSC \cite{wang2018path} & 0.1249 & 0.1653 & 0.1904 & 0.2955 & 0.1078 & 0.1183 & 0.1292 & 0.1485 \\
    & SPEM \cite{zhang2019inferring} & 0.1014 & 0.1113 & 0.1305 & 0.1787 & 0.0934 & 0.0965 & 0.1014 & 0.1103 \\
    & CLVA \cite{rakesh2019linked} & 0.1565 & 0.2201 & 0.3179 & 0.3868 & 0.1024 & 0.1397 & 0.1531 & 0.1754 \\
    \cline{2-10}
    & \textbf{A2CF}& \textbf{0.2187} & \textbf{0.2978} & \textbf{0.3899} & \textbf{0.5518} & \textbf{0.1532} & \textbf{0.1788} & \textbf{0.2021} & \textbf{0.2395} \\
    \hline
    \hline
        \multirow{5}{*}{Office} & Sceptre \cite{mcauley2015inferring} & 0.2489 & 0.3623 & 0.4762 & 0.6100 & 0.1570 & 0.1948 & 0.2242 & 0.2512 \\
    & PMSC \cite{wang2018path} & 0.1345 & 0.1924 & 0.2596 & 0.3701 & 0.0954 & 0.1182 & 0.1425 & 0.1759 \\
    & SPEM \cite{zhang2019inferring} & 0.0741 & 0.0906 & 0.1149 & 0.1528 & 0.0579 & 0.0632 & 0.0693 & 0.0735\\
    & CLVA \cite{rakesh2019linked} & 0.2322 & 0.3750 & 0.4869 & 0.6343 & 0.1480 & 0.1946 & 0.2378 & 0.2733 \\
    \cline{2-10}
    & \textbf{A2CF}& \textbf{0.3143} & \textbf{0.4220} & \textbf{0.5310} & \textbf{0.6644} & \textbf{0.2193} & \textbf{0.2540} & \textbf{0.2815} & \textbf{0.3101} \\
    \hline     
    \end{tabular}
\label{table:recommendation}
\vspace{-0.5cm}
\end{table*}

\textbf{Evaluating Recommendation Effectiveness.}  
We leverage two ranking metrics, namely hit ratio at rank $K$ (HR$@K$) and normalized discounted cumulative gain at rank $K$ (NDCG$@K$) that are widely adopted in recommendation research \cite{wang2020next,yin2015joint,wang2018streaming}. For each positive test instance $(u_i, v_q, v_{j^+})$, we mix the ground truth $v_{j^+}$ with $J$ random negative items, then rank all these $J+1$ items and compute the HR$@K$ and NDCG$@K$ scores. We set $J\!=\!1,000$ following \cite{chen2020sequence} and adopt $K=5,10,20,50$ for evaluation.

\textbf{Sampling Query Items.} 
However, a sticking point is the selection criteria of query item $v_q$. Though $v_{j^+}$ and $v_{j^-}$ can be easily obtained via each user's interaction record, the majority of popular e-commerce datasets like Amazon \cite{mcauley2015inferring} and Yelp \cite{chen2018pme} only record actual purchases as user-item interactions. Such limitation prevents us from knowing which other options (i.e., query items) that $u_i$ has browsed before deciding to purchase $v_{j^+}$. Hence, for each observed user-item interaction $(u_i, v_{j^+})$, we propose to leverage \textbf{\textit{popularity-biased sampling}} to pick $v_q$ to simulate the query item:
\begin{itemize}
	\item[I.] For $(u_i, v_{j^+})$, find all substitutes of $v_{j^+}$ that $u_i$ has never interacted with, denoted as $\mathcal{S}_{ij^+}$;
	\item[II.] For each $v_q \in \mathcal{S}_{ij^+}$, calculate the amount of users it has interacted with, denoted as $pop_q$;
	\item[III.] Draw one $v_q$ with probability $P(v_q) \sim pop_q^{0.75}$.
\end{itemize}
Note that there will be only one query item $v_q$ sampled for every $(u_i, v_{j^+})$ instance. As such, we can obtain triplets $(u_i, v_q, v_{j^+})$ as the ground truth for training and test. The rationale of popularity-biased sampling is that, when searching for a specific type of product to buy, customers usually tend to start with the most popular ones, and then expand their candidates as their demands become clearer. Motivated by successful graph theory practices \cite{chen2018pme,tang2015line}, we smoothen the item popularity with $0.75$ power. This moderately increases the chance of drawing unpopular items and helps validate the robustness of the model with diverse query compositions. It is worth noting that, in the test set, we restrict that the ground truth item $v_{j^+}$ in each instance $(u_i, v_q, v_{j^+})$ is neither paired with user $u_i$ nor $v_q$ for model training. This allows for generating the hardest possible test instances because the test item is always new for both the user and the query item.

\textbf{Evaluating Interpretation Quality.}
In the emerging area of explainable recommendation, the quality of text-based interpretations is mainly tested via either human evaluation \cite{zhang2014explicit,wang2018explainable} or visualization \cite{wang2019kgat,he2015trirank}. Despite being effective to some extent, they lack quantitative measurements for the generated interpretations. Because the process of generating attribute ranking lists for interpretation can be viewed as a special type of document retrieval task, metrics like mean average precision (MAP) from relevant domains \cite{li2010unified,yao2013automatic} can be used to evaluate the ranks of attributes mentioned in the actual user reviews. However, it is pointless to evaluate interpretations generated for non-ground truth items, as the user has never visited nor reviewed them. Therefore, we only consider the MAP score for the interpretation for each ground truth item $v_{j^+}$, and trade it off using the NDCG score for $v_{j^+}$. In this way, we define \textbf{\textit{attribute trade-off coverage}} (ATC) as a generic evaluation metric for attribute-aware interpretable recommendation models:
\vspace{-0.2cm}
\begin{equation}
	\!\!\! \textnormal{ATC} = \! \underset{\forall (u_i,v_q,v_{j^+})}{\textnormal{MEAN}} \frac {2\times \textnormal{MAP}_{iqj^+} \times \textnormal{NDCG}_{iqj^+}}{\textnormal{MAP}_{iqj^+} + \textnormal{NDCG}_{iqj^+}},
\vspace{-0.2cm}
\end{equation}
where MAP$_{iqj^+}$ is the mean average precision of $\{\Delta^{iqj}_n\}_{n=1}^{|\mathcal{A}|}$ (i.e., the attribute ranking list used for recommendation explanation) w.r.t. actually mentioned attributes in user $u_i$'s review on item $v_{j^+}$, and NDCG$_{iqj^+}$ is the NDCG score of ground truth item $v_{j^+}$ in the ranking list. By taking the harmonic mean of MAP$_{iqj^+}$ and NDCG$_{iqj^+}$, we expect a large ATC score only when the interpretation and recommendation are both accurate for test case $(u_i, v_q, v_{j^+})$. 

\vspace{-0.2cm}
\subsection{Baseline Methods}
We briefly introduce the baseline methods for comparison below.
\begin{itemize}
	\item \textbf{Sceptre}~\cite{mcauley2015inferring}\textbf{:} The key idea of Sceptre is combining topic modeling with supervised link prediction to predict possible substitutable relationships between two products.
	\item \textbf{PMSC}~\cite{wang2018path}\textbf{:} It firstly embeds products into relation-specific spaces, then incorporates multiple path constraints to enhance the expressiveness of learned product embeddings. 
	\item \textbf{SPEM}~\cite{zhang2019inferring}\textbf{:} It models substitutable product relationships by penalizing first-order proximity, rewarding second-order proximity and semantic similarity in the constructed product co-purchasing graph, thus making the representations of two substitutable products align closely in the latent space
	\item \textbf{CLVA}~\cite{rakesh2019linked}\textbf{:} As another state-of-the-art substitute recommendation model, it links two variational autoencoders conditioned on the observed links among items. On top of that, the integration of probabilistic matrix factorization further allows for personalized substitute recommendation.	
\end{itemize}

\vspace{-0.3cm}
\subsection{Hyperparameter Settings}\label{sec:hyperparamsetting}
To be consistent, we report the overall recommendation performance of A2CF with a unified set of parameters where $d=64$, $l=1$, $\gamma=0.7$ and $\beta\!=\! \epsilon\!=\!d^{0.5}\!=\!8$. The effect of different hyperparameter settings will be discussed in Section~\ref{sec:paramterimpact}. For all baselines, we use grid search to obtain their optimal hyperparameters.

\section{Experimental Results And Analysis}\label{sec:exp}
Following the settings in Section \ref{sec:exp_settings}, we conduct experiments\footnote{Public access to codes: https://bit.ly/bitbucket-A2CF} to evaluate the performance of A2CF regarding both recommendation effectiveness and interpretation quality. In particular, we aim to answer the following research questions (RQs) via experiments:
\begin{itemize}
	\item[\textbf{RQ1:}] How effectively can A2CF perform personalized substitute recommendation compared with state-of-the-art baselines?
	\item[\textbf{RQ2:}] How is the quality of attribute-based interpretations generated by A2CF?
	\item[\textbf{RQ3:}] What is the contribution of each key component of the proposed model structure?
	\item[\textbf{RQ4:}] How the hyperparameters affect the performance of A2CF in terms of recommendation effectiveness?
\end{itemize}

\begin{figure*}[t!]
\centering
\vspace{-0.4cm}
\includegraphics[width=6.8in]{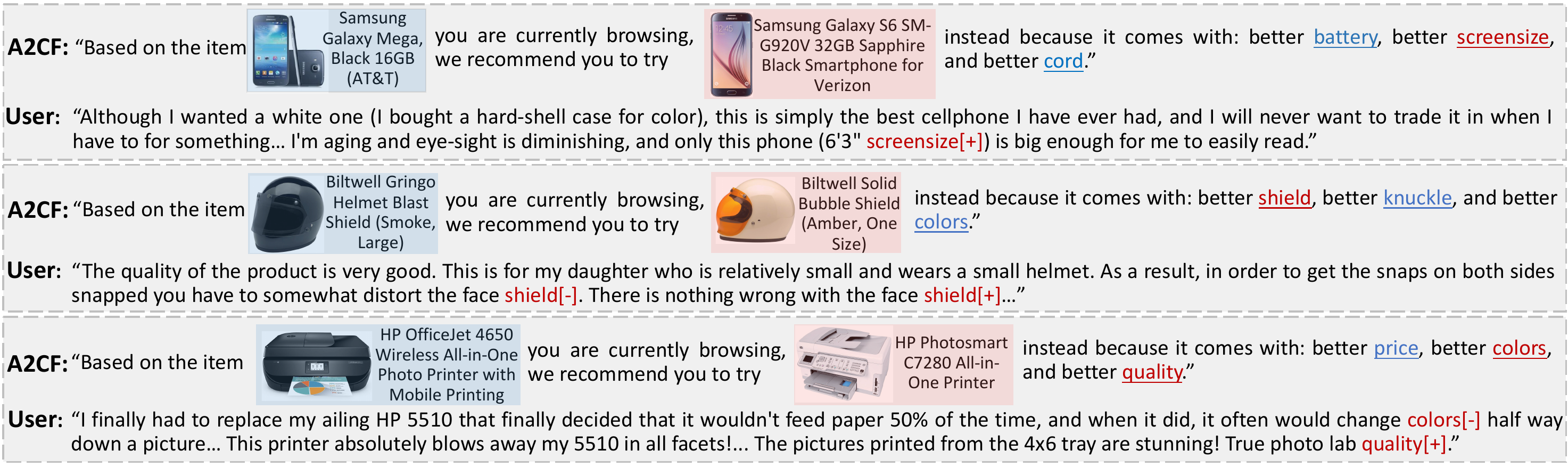}
\vspace{-0.4cm}
\caption{Case study on interpretations generated by A2CF on Cellphone (top), Automotive (middle), and Office (bottom). Comparisons are performed between interpretations and users' real reviews on the ground truth items. }
\label{fig:interpret}
\vspace{-0.5cm}
\end{figure*}

\vspace{-0.3cm}
\subsection{Recommendation Effectiveness (RQ1)}
We summarize all models' performance on personalized substitute recommendation with Table \ref{table:recommendation}. We discuss our findings as follows.

The first observation is that our proposed A2CF outperforms all baselines consistently and significantly ($p$-value $< 0.01$) on these three datasets. Compared with the second best results on all datasets, A2CF achieved an average improvement of 28.1\% and 29.2\% respectively on HR$@5$ and NDCG$@5$. As PMSC and SPEM do not take personalization into consideration, the recommended substitutes can hardly match users' personal preferences, leading to inferior performance. At the mean time, as a personalized substitute recommender, CLVA additionally utilizes latent user representations learned via probabilistic matrix factorization, and thus produces better recommendation results than PMSC and SPEM. However, by implementing substitution and personalization constraints via an attribute-aware scheme, A2CF represents the state-of-the-art effectiveness on this recommendation task.

Second, compared with deep learning models that are entirely based on product graphs (i.e., PMSC and SPEM), models that make use of the item attributes (i.e., A2CF, Sceptre and CLVA) show obvious advantages in terms of recommendation performance. This further validates the benefit of incorporating attribute information for personalized substitute recommendation. Particularly, though the pure LDA-based Sceptre is neither a personalized recommender nor enhanced by deep neural networks, it can still yield highly competitive results, and is even able to produce higher recommendation accuracy than CLVA in some cases (e.g., HR$@5$ on all three datasets). This is probably because when modelling items with attributes, Sceptre additionally introduces a category tree for selecting different attributes, which helps it learn more specific and high-quality attribute-based item representations.

Lastly, we find that the sparsity of user-item interactions and the amount of available attributes have opposite influences to the recommendation effectiveness of A2CF. On one hand, Cellphone and Office have similar sparsity (both are around 99.995\%), but the larger amount of attributes allows A2CF to thoroughly learn attribute-aware representations for both users and items, leading to better results on Office. One the other hand, despite offering the most attributes, the Automotive dataset witnesses relatively lower performance of A2CF due to its substantially higher sparsity.  

\vspace{-0.2cm}
\subsection{Interpretation Quality (RQ2)}
\textbf{Quantitative Analysis.} To quantitatively evaluate the interpretation quality of A2CF, we adopt the metric ATC proposed in Section \ref{sec:eva_protocol}, and compare A2CF with the other two attribute-based substitute recommenders Sceptre and CLVA. For both baselines, an item-attribute relevance score can be inferred, which enables us to obtain an attribute ranking list for each ground truth item to compute the ATC score. For fairness, we truncate the attribute ranking list of Sceptre, CLVA and A2CF with the length of 500 for evaluation because different numbers of attributes are used by them. Then, we present the quantitative results in Table \ref{table:interpretation}. Apparently, A2CF leads Sceptre and CLVA regarding ATC scores on all three datasets due to its ability to generate personalized interpretations.

\begin{table}[b]
\vspace{-0.8cm}
\caption{Quantitative results on interpretation quality.}
\vspace{-0.4cm}
\renewcommand{\arraystretch}{0.9}
\setlength\tabcolsep{8pt}
\center
  \begin{tabular}{|c|c|c|c|}
    \hline
    \multirow{2}{*}{Method} & \multicolumn{3}{c|}{ATC} \\
    \cline{2-4}
    & Cellphone & Automotive & Office\\
    \hline
    Sceptre & 0.0108 & 0.0094 & 0.0124 \\
	CLVA & 0.0096 & 0.0077 & 0.0112 \\
	\textbf{A2CF} & \textbf{0.0132} & \textbf{0.0113} & \textbf{0.0154} \\	
    \hline
\end{tabular}
\label{table:interpretation}
\end{table}

\textbf{Case Study.} We further conduct a case study in Figure \ref{fig:interpret} to qualitatively examine the interpretability of A2CF. Specifically, we visualize and compare the interpretations generated by A2CF and users' real review texts. Note that we use the top three attributes to explain each recommendation result, i.e., $Z=3$. In each case, the query item is marked blue, while the interpretation and review are for the same ground truth item in the red box. Each ground truth item is positioned higher than 10 in the ranking list. We use red to mark attributes that are mentioned in both the interpretation and the corresponding review, and label the sentiment of attributes (positive/negative) in the user review. For the first two users respectively browsing a cellphone and a helmet, A2CF successfully recognizes their demand on ``screensize'' and ``shield'', and recommends the correct items that perform better than the query items on corresponding attributes. We can also see there is positive user feedback on both the ``screensize'' and ``shield'' of the recommended cellphone and helmet. Furthermore, in the third case, two better attributes ``colors'' and ``quality'' of the recommended printer are used for interpretation, which are both mentioned by the user. Though attribute ``colors'' expresses negative sentiment in this review, in fact the user is complaining about a previously purchased printer. This further validates our model's capability of aligning item attributes with user preferences to generate interpretations. 

\vspace{-0.7cm}
\subsection{Importance of Key Components (RQ3)}\label{sec:ablation}
To better understand the performance gain from the major components proposed in A2CF, we perform ablation analysis on different degraded versions of A2CF. Table \ref{table:ablation} summarizes the recommendation outcomes in terms of HR$@$10. 

\begin{figure*}[t!]
\centering
\vspace{-0.3cm}
\begin{tabular}{cccc}
	\vspace{-0.2cm}\includegraphics[width=1.75in]{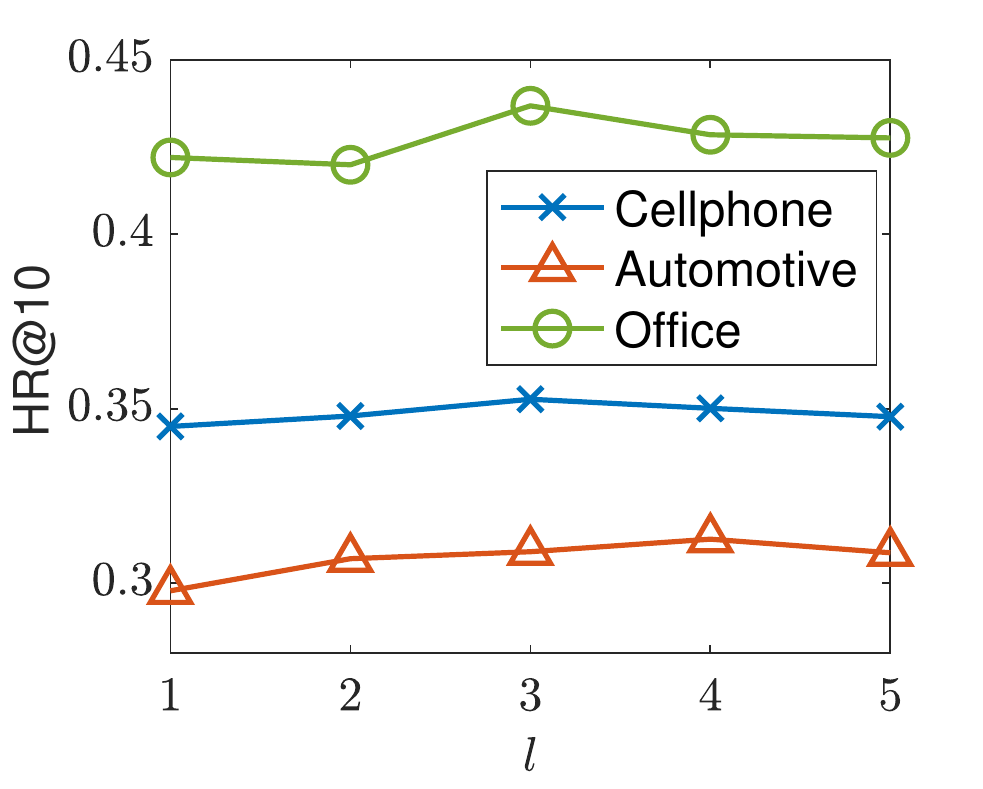}
	&\hspace{-0.6cm}\includegraphics[width=1.75in]{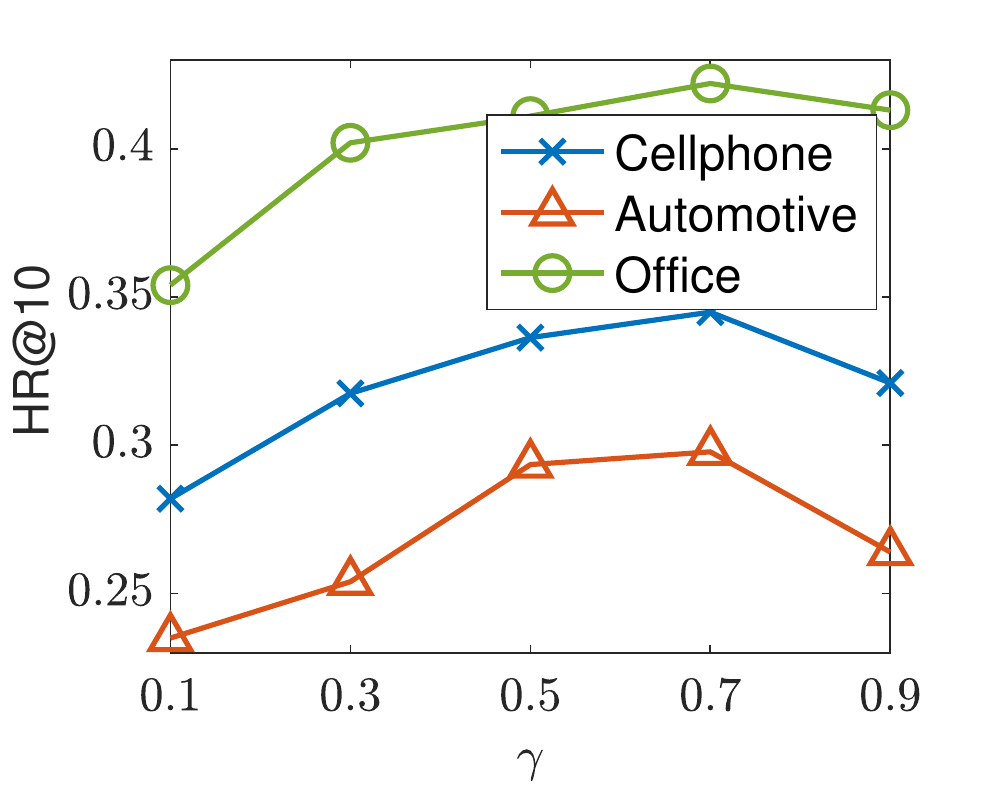}
	&\hspace{-0.6cm}\includegraphics[width=1.75in]{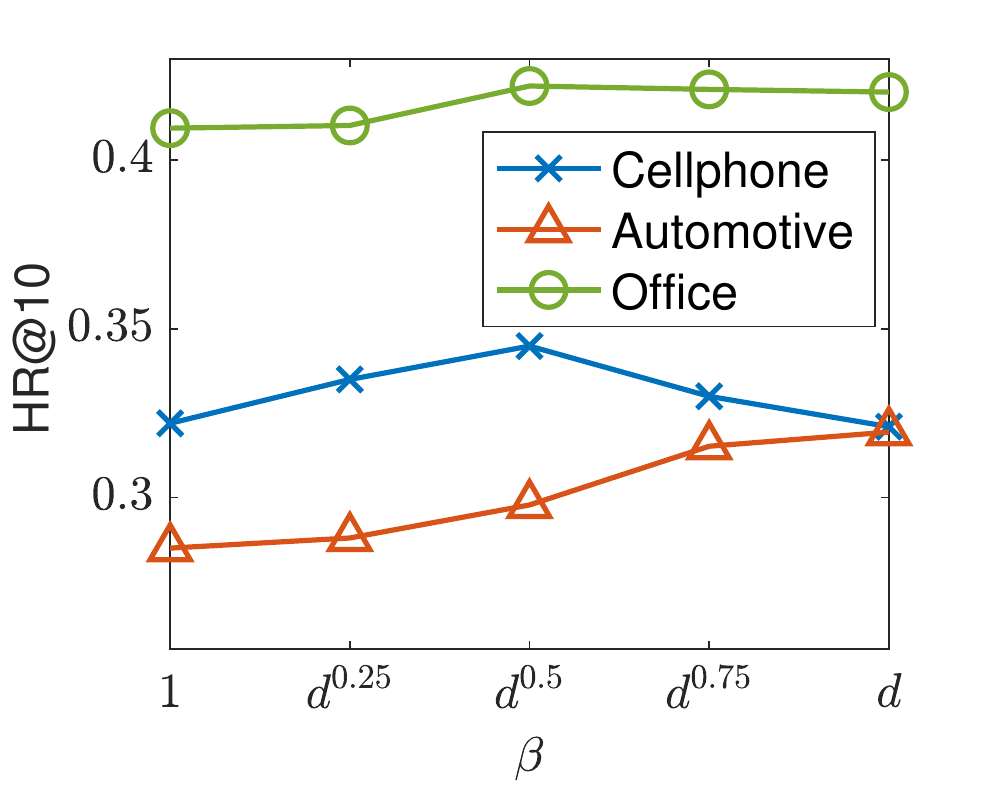}
	&\hspace{-0.6cm}\includegraphics[width=1.75in]{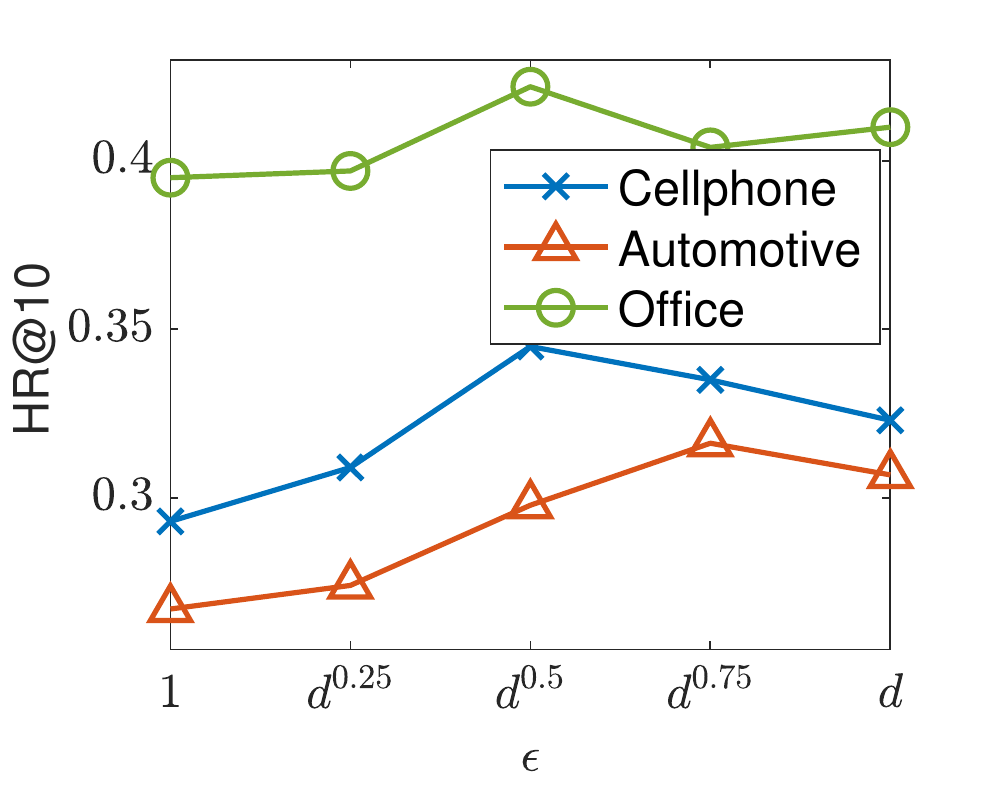}\\
		\end{tabular}
\vspace{-0.4cm}
\caption{Recommendation performance w.r.t. $l$, $\gamma$, $\beta$ and $\epsilon$.}
\label{Figure:paramsensitivity}
\vspace{-0.5cm}
\end{figure*}

\textbf{Removing Substitution Constraint.}
The substitution constraint in Eq.(\ref{eq:BPR-S-rewrite}) enforces that the recommended items should hold substitutable relationship with the query item. We remove the substitution constraint by setting $\gamma = 0$. After that, a severe performance decrease can be observed from all datasets, where the performance on Automotive is the most vulnerable. Because Automotive contains substantially more items than Cellphone and Office, the substitution constraint is of great significance to helping A2CF filter out irrelevant items and ensure successful recommendations. 

\textbf{Removing Personalization Constraint.}
Similar to the previous variant, we remove the personalization constraint by setting $\gamma = 1$ in Eq.(\ref{eq:BPR-S-rewrite}). As personal preferences are not properly captured in this variant, A2CF suffers from a significant performance drop due to the absence of personalization constraint. Compared with removing substitution constraint, the deletion of personalization constraint has more negative effect on datasets with relatively denser user-item interactions (i.e., Cellphone and Office). 

\begin{table}[t]
\caption{Ablation test with different model architectures.}
\vspace{-0.4cm}
\centering
\renewcommand{\arraystretch}{0.8}
\setlength\tabcolsep{0.3pt}
  \begin{tabular}{|c|c|c|c|}
    \hline
     \multirow{3}{*}{Variant} & \multicolumn{3}{c|}{HR$@$10}\\
    \cline{2-4}
    & Cell- & Auto- & \multirow{2}{*}{Office}\\
    & phone & motive & \\
 	\hline
    Default & \textbf{0.3449} & \textbf{0.2978} & \textbf{0.4220} \\
    \hline
    \makecell{Removing Substitution Constraint \\ \small{Eq.(\ref{eq:BPR-S-rewrite}) $\rightarrow - \!\!\!\!\!\!\!\!\!\!\!\!\!\!\!\!\!\!\!\! \sum\limits_{(u_i, v_{j^+}, v_{j^-})\in \mathcal{D}_{Rec}} \!\!\!\!\!\!\!\!\!\!\!\!\!\! \sigma' \Big{(} f_P(u_i,\! v_{j^+}\!) - f_P(u_i,\! {v_{j^-}}) \Big{)}$}} & 0.2772 & 0.1997 & 0.3745 \\
   	\hline
     \makecell{Removing Personalization Constraint \\ \small{Eq.(\ref{eq:BPR-S-rewrite}) $\rightarrow - \!\!\!\!\!\!\!\!\!\!\!\!\!\!\!\!\!\!\!\!\!\! \sum\limits_{(v_q, v_{j^+}, v_{j^-})\in \mathcal{D}_{Rec}} \!\!\!\!\!\!\!\!\!\!\!\!\!\! \sigma' \Big{(}f_S(v_q, \!v_{j^+}\!) - f_S(v_q,\! v_{j^-}) \Big{)}$}} & 0.2523 & 0.2312 & 0.3510 \\
    \hline
     \makecell{Removing Attribute Aggregation for Items \\ \small{Eq.(\ref{eq:projection}) $\rightarrow \textbf{w}_s^{\top}[\textbf{v}_q \odot \textbf{v}_j]$}} & 0.3116 & 0.2733 & 0.3994 \\
    \hline
     \makecell{Removing Attribute Aggregation for Users \\ \small{Eq.(\ref{eq:projection2}) $\rightarrow \textbf{w}_p^{\top}[\textbf{u}_i \odot \textbf{v}_j]$ }} & 0.3075 & 0.2708 & 0.3763 \\
     \hline
    \end{tabular}
\label{table:ablation}
\vspace{-0.8cm}
\end{table}

\textbf{Removing Attribute Aggregation for Items.}
With the removal of the attribute aggregation for items in Eq.(\ref{eq:projection}), there is a noticeable performance decrease for A2CF. The aggregated attribute-aware item representation (i.e., $\widetilde{\textbf{v}}_j$) aims to fully leverage the attributes to offer more information for item-item substitutable relationship predictions. Hence, this verifies the efficacy of merging attribute information with latent item representations for identifying substitutable relationships among items.

\textbf{Removing Attribute Aggregation for Users.}
We build the forth variant of A2CF by deleting the aggregated attribute information for users in Eq.(\ref{eq:projection2}). Generally, the inferior performance on three datasets indicates that considering users' preferred attributes practically improves user-item preference modelling. 

\vspace{-0.2cm}
\subsection{Impact of Hyperparameters (RQ4)}\label{sec:paramterimpact}
We answer RQ4 by investigating the performance fluctuations of A2CF with varied hyperparameters. Particularly, we study our model's sensitivity to network depth $l$, coefficient $\gamma$, as well as two scaling factors $\beta$ and $\epsilon$. Based on the standard hyperparameter setup in Section \ref{sec:hyperparamsetting}, we vary the value of one hyperparameter while keeping the others unchanged, and record the new recommendation result achieved. 
Similar to Section~\ref{sec:ablation}, HR$@$10 is used for demonstration. Figure \ref{Figure:paramsensitivity} lays out the results with different parameter settings.

\textbf{Impact of $l$.}
The value of the network depth $l$ is examined in $\{1, 2, 3, 4, 5\}$. In general, A2CF benefits from a relatively larger $l$ on all three datsets, but the performance improvement tends to stop when $l$ reaches a certain size (3 and 4 in our case) due to overfitting. 

\textbf{Impact of $\gamma$.}
We also study the impact of $\gamma \in \{0.1, 0.3, 0.5, 0.7, 0.9\}$ which balances the substitution and personalization constraints. The HR$@10$ scores on all datasets imply that, in A2CF, by laying enough emphasis on the substitution constraint ($0.7$ in our case), the joint effect of personalization and substitution can be maximized for recommendation. 

\textbf{Impact of $\beta$ and $\epsilon$.} 
The scaling factors $\beta$ and $\epsilon$ are respectively introduced in Eq.(\ref{eq:softmax1}) and Eq.(\ref{eq:softmax2}) to control the influence of dominating attributes when aggregating attribute information for items and users. As can be concluded from Figure \ref{Figure:paramsensitivity}, A2CF behaves differently on varied datasets when the scaling factors are adjusted in $\{1,d^{0.25},d^{0.5},d^{0.75},d\}$ ($d=64$ in our setting). 
This is possibly caused by the differences in the quantity of attributes extracted from three datasets. Since Automotive has more attributes due to higher diversity of items, larger $\beta$ and $\epsilon$ will encourage a more scattered probability distribution over all attributes, so more attributes are taken into consideration when modelling users and items, making the model adaptive to high data sparsity.

\vspace{-0.2cm}
\section{Related Work}\label{sec:related}
\subsection{Product Relationship Mining}
Understanding how products relate to each other is important to the fulfilment of customer satisfaction in different online shopping stages. In economics literature, two common product relationships are substitution and complement \cite{mas1995microeconomic}. Recently, product relationship mining has become a prospective research direction to enhance existing recommender systems that are unable to differentiate the relationships among products. The problem of mining product relationships is first introduced to recommendation research in \cite{mcauley2015inferring}, where a topic model-based approach Sceptre is proposed. With latent Dirichlet allocation (LDA), Sceptre infers product relationships by comparing the topic distributions of two products. More recently, the trend of utilizing reviews has carried over to neural network approaches, such as RRN \cite{zhang2019identifying} that learns product embeddings from both the textual reviews and manually crafted features. In contrast to those hybrid models combining reviews with product graphs, pure graph-based methods like PMSC \cite{wang2018path} and SPEM \cite{zhang2019inferring} are also introduced to fully utilize various properties in a product graph, e.g., path constraints and node proximities. 

Unfortunately, as recommender systems, all the aforementioned methods treat substitute recommendation as a pure item-based search task, thus neglecting the key aspect of personalization. As a recent solution, CLVA \cite{rakesh2019linked} fuses probabilistic matrix factorization with variational autoencoders to support personalized recommendation by jointly modelling user ratings leverages and item relationships. However, like most existing product relationship mining methods, the latent factor-based representation learning scheme in CLVA hinders it from offering explainable recommendations, rendering it difficult to fully convince and attract potential customers. 

\vspace{-0.2cm}
\subsection{Attribute Modelling for Recommendation}
In conventional top-$K$ recommendation, a large body of works adopt collaborative filtering, e.g., matrix factorization to utilize implicit interactions to infer the links between users and items \cite{koren2009matrix,rendle2009bpr,liu2019diverse}. The predominant problem is usually learning latent factors for users and items, and predicting user-item interactions by ranking the inner product of user and item embeddings \cite{koren2009matrix}.

Despite the promising recommendation quality of latent factor models, their latent characteristics make it frustratingly difficult to explain the generated recommendations \cite{wang2018explainable}, and such methods fail to adaptively generate recommendations when we are aware of a user's most cared features \cite{zhang2014explicit}. To overcome the limitations of latent factor models, review-based attribute modelling has attracted growing attention in recommendation research \cite{cheng2018aspect,cheng20183ncf}. Attributes are commonly extracted from reviews \cite{guan2019attentive,he2015trirank,cheng2018aspect} or tags \cite{yin2016adapting,yin2016discovering,wang2015geo} associated with each item. Such methods assume that a user's decision of purchase is largely determined by how each product meets her/his expectations on several key attributes \cite{zhang2014explicit}. Ganu et al. \cite{ganu2013improving} utilize aspects and sentiments manually crafted from restaurant reviews to enhance the performance of rating prediction. With the increasingly available review data, Zhang et al. \cite{zhang2014explicit} successfully combine techniques for automated attribute-level sentiment analysis with matrix factorization. Other subsequent variants are proposed to better capture the complex yet subtle relationships among users, products and attributes, such as modelling such heterogeneous relationships as a tripartite graph with TriRank \cite{he2015trirank}, utilizing joint factorization in MTER \cite{wang2018explainable}, and incorporating attention networks into AARM \cite{guan2019attentive}. Attribute modelling opens up an opportunity for generating interpretations for the recommended items. Existing attribute-aware methods, however, can only produce generic explanations \cite{wang2018explainable} as they do not account for the attention shifts among different attributes when users are browsing different items. In contrast, our proposed A2CF is able to firstly infer a user's desired attributes, and retrieve advantageous attributes of the recommended items to construct personalized interpretations.

\vspace{-0.2cm}
\section{Conclusion}\label{sec:conclusion}
In this paper, we study a new research problem, namely personalized and interpretable substitute recommendation, and propose a novel  A2CF model as a solution. A2CF fully utilizes attribute information extracted from reviews, which effectively bridges user preferences and item properties for generating personalized substitute recommendations and endows the model with high explainability. The experimental results evidence that A2CF can yield superior performance on both recommendation and interpretation.

\section*{Acknowledgement}
This work has been supported by Australian Research Council (Grant No. DP190101985, DP170103954 and FT200100825) and National Natural Science Foundation of China (Grant No. NSFC 61806035, U1936217, 61732008 and 61725203).

\bibliographystyle{ACM-Reference-Format}

\end{document}